\documentclass[aps,pra,twocolumn,showpacs]{revtex4}
\usepackage[dvips]{graphicx}
\usepackage{amsmath}
\usepackage{times}
\usepackage{psfrag}

\begin{document}

\title
{
Spin-1 bosons with coupled ground states in optical lattices
}

\author
{K.~V.~Krutitsky
and R.~Graham
}

\affiliation{
Fachbereich Physik der Universit\"at Duisburg-Essen, Campus Essen,
Universit\"atsstr. 5, 45117 Essen, Germany
}

\date{\today}

\begin{abstract}
The superfluid--Mott-insulator phase transition of ultracold
spin-1 bosons with ferromagnetic and antiferromagnetic
interactions in an optical lattice is theoretically investigated.
Two counterpropagating linearly polarized laser beams with the
angle $\theta$ between the polarization vectors (lin-$\theta$-lin
configuration), driving an $F_g=1$ to $F_e=1$ internal atomic
transition, create the optical lattice and at the same time couple
atomic ground states with magnetic quantum numbers $m=\pm 1$. Due
to the coupling the system can be described as a two-component
one. At $\theta=0$ the system has a continuous isospin symmetry,
which can be spontaneously broken, thereby fixing the number of
particles in the atomic components. The phase diagram of the
system and the spectrum of collective excitations, which are
density waves and isospin waves, are worked out. In the case of
ferromagnetic interactions, the superfluid--Mott-insulator phase
transition is always second order, but in the case of
antiferromagnetic interactions for some values of system
parameters it is first order and the superfluid and Mott phases
can coexist. Varying the angle $\theta$ one can control the
populations of atomic components and continuously turn on and tune
their asymmetry.
\end{abstract}

\pacs{03.75.Lm,03.75.Mn,71.35.Lk}

\maketitle

\section{Introduction}

The investigation of quantum phase transitions (QPT's) in optical
lattices has become one of the major subjects in the physics of
ultracold atoms, especially after a convincing experimental
realization in the system of spin-polarized
$^{87}$Rb~\cite{Greiner}. One of the interesting directions in
this rapidly developing field is the study of QPT's in ultracold
bosonic atoms with spin degrees of freedom. Systematic studies of
QPT's in spin-1 bosonic systems with antiferromagnetic
interactions have been performed very recently. It has been shown
that there are different Mott insulating phases such as spin
singlet, nematic, and dimerized ones~\cite{DZ,Yip,Zhou,ILD,SZ},
and the mean-field analysis shows that the transition between the
singlet and nematic phases is weakly first order~\cite{ILD,SZ}.
Fragmented condensates with topological excitations have been
studied in Ref.~\cite{DZ}. In the presence of a uniform magnetic
field, different polar superfluids with spins aligned along or
opposite to the field direction are possible~\cite{SC}. QPT's of
spin-2 bosons have been studied as well~\cite{HG}.

In the present paper, we shall investigate QPT's of \mbox{spin-1}
bosons with coupled ground states in an optical lattice and their
collective excitations. We consider a generalization of the setup
we have studied recently~\cite{KG03}, where the population of only
two degenerate ground states with magnetic quantum numbers $m=\pm
1$ was considered. The presence of atoms in all the three ground
states leaves two low-lying linear combinations, which transform
in the fundamental representation of an isospin SU(2) which
determines the symmetry and the dynamics of the resulting system.
Varying the laser polarization one can continuously turn on and
tune the asymmetry between the components and control the
populations of components in the superfluid and the Mott phases.
As we shall show, a rather rich scenario of quantum phase
transitions of first and/or second order becomes accessible to
experiments in this way. In the present paper, we explore the
phase diagram and the low-lying excitations of the various phases.
In Section II we introduce the laser configuration and the
Hamiltonian which describes the system. The Bose-Hubbard model for
this system is obtained in Section III, and its isospin symmetry
is discussed. Section IV gives a derivation of the phase diagram
in the mean-field approximation and contains our results on
possible coexistence of Mott and superfluid phases and first- and
second-order phase transitions. In Section V the collective
excitations consisting of Bogoliubov mode and isospin waves with
or without a gap are analyzed. The final section summarizes our
conclusions.

\section{System}

We consider a system of neutral polarizable spin-1 bosonic atoms
of mass $M$, possessing three Zeeman-degenerate internal ground
and excited electronic states characterized by the magnetic
quantum number $m = 0,\pm 1$ ($F_g=F_e=1$), in a (quasi-)
one-dimensional optical lattice. The latter is assumed to be
created by two counterpropagating linearly polarized laser waves
of equal amplitudes and frequencies with the wave number $k_L$,
and the angle $\theta$ (with $0\le\theta\le\pi/2$) between the
polarization vectors (lin-$\theta$-lin configuration)~\cite{LAL},
and with detuning  $\Delta$ from the internal atomic transition.
The running laser waves form left- and right- polarized standing
waves with the Rabi frequencies
\begin{equation}
\Omega_\pm(z)
=
\tilde\Omega_0
\cos( k_L z \pm \theta/2)
\;,
\end{equation}
which couple internal ground and excited states by $V$ and
$\Lambda$ transitions. In order to avoid decoherence due to
spontaneous emission, $|\Delta|$ must be much larger than the
spontaneous emission rate. Because of the large detuning the
excited state can be adiabatically eliminated. The resulting
effective Hamiltonian for the atoms in the three degenerate ground
states has the following form~\cite{Ho,OM}:
\begin{eqnarray}
\label{H}
H
&=&
\int
\,dz
\left(
\frac{\hbar^2}{2M}
\frac{\partial\hat\psi^\dagger_\alpha}{\partial z}
\frac{\partial\hat\psi_\alpha}{\partial z}
+
\frac{\hbar}{\Delta}
\hat\psi^\dagger_\alpha
\Omega^2_{\alpha\beta}
\hat\psi_\beta
\right.
\\
&+&
\left.
\frac{g_s}{2}
\hat\psi_\alpha^\dagger
\hat\psi_\beta^\dagger
\hat\psi_\beta
\hat\psi_\alpha
+
\frac{g_a}{2}
\hat\psi_{\alpha}^\dagger
\hat\psi_{\alpha'}^\dagger
{\bf F}_{\alpha\beta}
\cdot
{\bf F}_{\alpha'\beta'}
\hat\psi_{\beta'}
\hat\psi_{\beta}
\right)
\,,
\nonumber
\end{eqnarray}
where $\hat\psi_\alpha(z)$ is the bosonic field annihilation
operator for the atom in the hyperfine ground state
$|F=1,\alpha\rangle$. Summation over repeated spin indices
$\alpha, \beta$ is implied. ${\bf F}_{\alpha\beta}$ is the vector
of $F=1$ $(3\times 3)$ matrices in the representation furnished by
the three degenerate ground states. The matrix
\begin{equation}
\Omega^2
=
\left(
    \begin{array}{ccc}
       \Omega_+^2        &  0          & \Omega_+ \Omega_-\\
       0                 &  \Omega_0^2 & 0\\
       \Omega_+ \Omega_- &  0          & \Omega_-^2
    \end{array}
\right)
\:,
\end{equation}
where
\begin{equation}
\label{omega0}
\Omega_0^2
=
\Omega_+^2+\Omega_-^2
=
\tilde\Omega_0^2
\left(
    1
    +
    \cos\theta
    \cos 2 k_L z
\right)
\;,
\end{equation}
determines the lattice potential with the period $\pi/k_L$ and at
the same time couples the atomic ground states with $m = \pm 1$.
The parameters $g_{s,a}$ describe the repulsive interaction of the
condensate atoms and the spin-changing collisions. In a (quasi-)
one-dimensional case they have the form~\cite{1DBEC} $ g_{s,a} =
{4 \hbar^2 a_{s,a}}/{M a_\perp^2} $, where $a_{s,a}$ are symmetric
and antisymmetric scattering lengths, and
$a_\perp=\sqrt{2\hbar/M\omega_\perp}$ is the size of the ground
state for the harmonic potential with the frequency $\omega_\perp$
confining the atomic system in the transverse directions. In our
previous work~\cite{KG03}, we have studied the case when the
ground state with $m=0$ is not occupied. In the present paper, we
shall deal with the situations when this state is occupied as
well.

\section{Bose-Hubbard model}

The $V$ and $\Lambda$ laser-induced transitions lead to two sets
of orthogonal Bloch eigenmodes which we denote by the indices $0$
and $\Lambda$, respectively. We assume that the atoms stay always
in the lowest Bloch bands with the dispersion relations $E_0(k)$
and $E_\Lambda(k)$. As explained below, in the present work we are
interested in the case of $\Delta<0$ and $\theta \approx 0$. In
this special case the lowest-band approximation is well justified,
because the lowest Bloch bands are separated from the first
excited ones by an energy of the order of
$\sqrt{E_R\hbar\tilde{\Omega}_0^2/|\Delta|}$, where
$E_R=\hbar\omega_R$ is the recoil energy. Then the spinor-field
operator $\hat{\bf\Psi}$ can be decomposed in the Wannier basis
localized at the minima of the lattice potential labelled by $i$,
\begin{equation}
\label{Psi}
\hat{\bf\Psi}(z)=
\sum_{i}
\sum_{\sigma=0,\Lambda}
\exp
\left(
   i \varphi_{\sigma i}
\right)
{\bf W}_{\sigma i}(z)
\hat a_{\sigma i}
\,,
\end{equation}
where ${\bf W}_{\sigma i}(z)={\bf W}_\sigma(z-z_i)$ are
three-component Wannier spinors for the lowest energy bands. They
are obtained by the solution of the eigenvalue problem for the
Hamiltonian (\ref{H}) in the case $g_{s,a}=0$ and satisfy the
orthonormality condition $\int {\bf W}_{\sigma i}^\dagger(z)\cdot
{\bf W}_{\sigma' j}(z) \,dz = \delta_{ij} \delta_{\sigma\sigma'}
$. They have the form
\begin{equation}
\label{Wannier}
{\bf W}_{0 i}=(0,W_{0i},0)^T, \quad {\bf W}_{\Lambda i}=(W_{+i},0,W_{-i})^T.
\end{equation}
The indices $i,j$ label the sites of the one-dimensional periodic lattice.
The $a_{\sigma i}$ is the Bose annihilation operator attached to the $i$th lattice site.
The phases $\varphi_{\sigma i}$ are not yet defined and their proper choice is
discussed in a moment. The hyperfine spin variable at lattice site $i$ is given by
\begin{equation}
\hat{{\bf F}}_i
=
\sum_{\sigma\sigma'}
\hat{a}^\dagger_{\sigma i}
{\bf F}_{\sigma\sigma' i}
\hat{a}_{\sigma' i}
e^{-i(\varphi_{\sigma i}-\varphi_{\sigma' i})}
\end{equation}
with
\begin{equation}
{\bf F}_{\sigma\sigma' i}
=
\int dz {\bf W}^\dagger_{\sigma i}(z)
\cdot{\bf F}_i
\cdot{\bf W}_{\sigma' i}(z)
\;.
\end{equation}

Substituting Eq.~(\ref{Psi}) into Eq.~(\ref{H}) and taking into
account only the hopping between the nearest lattice sites and the
on-site atomic interactions, we obtain the two-component
Bose-Hubbard Hamiltonian
\begin{eqnarray}
\label{BHH}
&&
\hat H_{BH}
=
-\sum_{\sigma}
\left|
    J_\sigma
\right|
\sum_{<i,j>}
\hat a_{\sigma i}^\dagger
\hat a_{\sigma j}
\nonumber\\
&&
+
\sum_{\sigma}
\frac{U_\sigma}{2}
\sum_i
\hat n_{\sigma i} (\hat n_{\sigma i}-1)
+
K
\sum_i
\hat n_{0i} \hat n_{\Lambda i}
\nonumber\\
&&
-\frac{|P|}{2}
\sum_i
\left(
    \hat a_{0i}^\dagger
    \hat a_{0i}^\dagger
    \hat a_{\Lambda i}
    \hat a_{\Lambda i}
    +
    \hat a_{\Lambda i}^\dagger
    \hat a_{\Lambda i}^\dagger
    \hat a_{0i}
    \hat a_{0i}
\right)
\nonumber\\
&&
-
\delta
\sum_i
\hat n_{0i}
-\mu \sum_{\sigma,i} \hat n_{\sigma i}
\,,
\end{eqnarray}
where $\mu$ is a chemical potential.
The tunneling matrix elements
\begin{equation}
J_\sigma
=
-\int
{\bf W}_{\sigma,i+1}^\dagger(z)\cdot
\left(
-
\frac{\hbar^2}{2M}
\frac{\partial^2}{\partial z^2}
+
\frac{\hbar}{\Delta}
\hat\Omega^2
\right)\cdot
{\bf W}_{\sigma,i}(z)
\,dz
\;,
\end{equation}
the atomic interaction parameters
\begin{eqnarray}
U_\Lambda
&=&
\int
\left[
    \left(
        g_s + g_a
    \right)
    \left(
        \left|
            W_{+i}
    \right|^2
    +
        \left|
            W_{-i}
    \right|^2
    \right)^2
\right.
\nonumber\\
&&
\left.
    -
    4 g_a
    \left|
        W_{+i}
    \right|^2
    \left|
        W_{-i}
    \right|^2
\right]
\,dz
\;,
\nonumber\\
U_0
&=&
g_s
\int
\left|
    W_{0i}
\right|^4
\,dz
\;,
\\
K
&=&
\left(
    g_s + g_a
\right)
\int
\left|
    W_{0i}
\right|^2
\left(
    \left|
        W_{+i}
    \right|^2
    +
    \left|
        W_{-i}
    \right|^2
\right)
\,dz
\;,
\nonumber\\
P
&=&
2 g_a
\int
\left(
    W_{0i}^*
\right)^2
W_{+i}
W_{-i}
\,dz
\;,
\nonumber
\end{eqnarray}
and the relative shift of the mean energies of the eigenmodes
\begin{eqnarray}
\delta
&=&
\int
{\bf W}_{\Lambda,i}^\dagger(z)\cdot
\left(
-
\frac{\hbar^2}{2M}
\frac{\partial^2}{\partial z^2}
+
\frac{\hbar}{\Delta}
\hat\Omega^2
\right)\cdot
{\bf W}_{\Lambda,i}(z)
\,dz
\nonumber\\
&-&
\int
{\bf W}_{0,i}^\dagger(z)\cdot
\left(
-
\frac{\hbar^2}{2M}
\frac{\partial^2}{\partial z^2}
+
\frac{\hbar}{\Delta}
\hat\Omega^2
\right)\cdot
{\bf W}_{0,i}(z)
\,dz
\nonumber\\
&=&
\frac{1}{k_L}
\int_0^{k_L}
\left[
    E_\Lambda(k)
    -
    E_0(k)
\right]
\,dk
\end{eqnarray}
can be simultaneously changed by varying the laser intensity
(which is proportional to $\tilde{\Omega}_0^2$)
and/or the angle $\theta$,
but the variations of $J_\sigma$ and $\delta$ are much faster.
The parameter $P$ can be either positive or negative depending
on the sign of the antisymmetric coupling $g_a$.

The phases $\varphi_{\sigma i}$ are determined from the
requirement of minimal energy of the Bose-Hubbard Hamiltonian,
which amounts to demanding that $J_\sigma\cos(\varphi_{\sigma
i}-\varphi_{\sigma,i+1})$ and $P\cos(2\varphi_{0
i}-2\varphi_{\Lambda i})$ are maximal. This requirement leading to
\begin{eqnarray}
\varphi_{\sigma,i+1}
=
\left\{
    \begin{array}{ll}
       \varphi_{\sigma i}         & \textrm{if $J_\sigma>0$}\\
       \varphi_{\sigma i} \pm \pi & \textrm{if $J_\sigma<0$}
    \end{array}
\right.\label{9}
\\
\varphi_{0i}
=
\left\{
    \begin{array}{ll}
       \varphi_{\Lambda i}           & \textrm{if $g_a<0$}\\
       \varphi_{\Lambda i} \pm \pi/2 & \textrm{if $g_a>0$}
    \end{array}
\right.
\end{eqnarray}
has been taken into account in the derivation of the Hamiltonian
(\ref{BHH}), which therefore only depends on the absolute values
$|J_\sigma|$ and $|P|$ besides the other system parameters
$U_\sigma$, which are positve in the examples we consider and of
about equal size, $K$, which can be larger or smaller than the
$U_\sigma$ depending on the sign of the antisymmetric coupling
$g_a$, and $\delta$. As one can clearly see from
Eq.~(\ref{omega0}) $J_0$ depends on $\tilde\Omega_0^2 \cos
\theta/|\Delta|$ and it is always positive~\cite{Jaksch}. In
principle $J_\Lambda$ can also become negative for large enough
$\theta$~\cite{KG03}, but at such values of $\theta$ the shift
$\delta$ becomes comparable to the atomic interaction parameters
in Hamiltonian (\ref{BHH}) and it becomes energetically more
favorable for atoms to stay in the $0$ mode. In the present work,
we are interested in the situations when both modes are occupied.
Therefore we have to restrict ourselves to small values of
$\theta$ and consider $J_\Lambda$ as well as $J_0$ as positive
quantities. Thus in the following always the first of the
alternatives in Eq.~(\ref{9}) applies. In general, $J_0 \ge
J_\Lambda$.

In the present paper, we investigate only the case of red detuning
($\Delta<0$), since in this case the minima of the effective
periodic potentials for the $\Lambda$ and $0$ modes coincide which
leads to higher values of $K$ and $|P|$ compared to the case of
blue detuning ($\Delta>0$). In addition if $\theta=0$,
$W_{+i}=W_{-i}=W_{0i}/\sqrt{2}$ and we have
\begin{equation}
\label{simpl}
J_0=J_\Lambda=J
\;,\;
U_0=U_\Lambda=U
\;,\;
K=U+P
\;,\;
\delta=0
\;.
\end{equation}
The $0$ and $\Lambda$ modes are degenerate in this case, because
$E_0(k)=E_\Lambda(k)$ for any value of the laser intensity. Indeed
the Hamiltonian (\ref{BHH}) then has an enhanced U(1)$\times$U(1)
symmetry. This can be seen by defining the global generators of an
SU(2) algebra (with $L$ the number of lattice sites)
\begin{eqnarray}
\hat{T}_1
&=&
\frac{1}{2L}
\sum_{i=1}^L
  \left(\hat{a}_{\Lambda i}^\dagger
  \hat{a}_{0 i}
  +
  \hat{a}_{0 i}^\dagger
  \hat{a}_{\Lambda i}\right)
\:,
\nonumber\\
\hat{T}_2
&=&
\frac{i}{2L}
\sum_{i=1}^L
  \left(\hat{a}_{\Lambda i}^\dagger
  \hat{a}_{0 i}
  -
  \hat{a}_{0 i}^\dagger
  \hat{a}_{\Lambda i}\right)
\;,
\\
\hat{T}_3
&=&
\frac{1}{2L}
\sum_{i=1}^L
\left(
    \hat{a}_{0 i}^\dagger
    \hat{a}_{0 i}
    -
    \hat{a}_{\Lambda i}^\dagger
    \hat{a}_{\Lambda i}
\right)
\;.
\nonumber
\end{eqnarray}
In terms of these generators the density of the hyperfine spin
variable $\hat{{\bf f}}=\sum_i\hat{{\bf F}}_i/L$ for $\theta=0$,
which always points in the $x$ direction defined by the direction
of the linear polarization of the external laser, is given by
$\hat{{\bf f}}=2\hat{T}_{1(2)}{\bf e}_x/L$ for $g_a<0$ ($g_a>0$).

\begin{figure}[t]
\centering

\psfrag{U}[cc]{\rotatebox{180}{$10^{3}\,U_\Lambda/E_R$, $10^{3}\,U_0/E_R$}}
\psfrag{t}[tc]{$\theta$ (deg)}
\psfrag{P}[br]{$|P|$}
\psfrag{UL}[bl]{$U_\Lambda$}
\psfrag{U0}[bc]{$U_0$}
\psfrag{p}[bc]{\rotatebox{180}{$10^{5}\,|P|/E_R$}}

  \includegraphics[width=5.5cm]{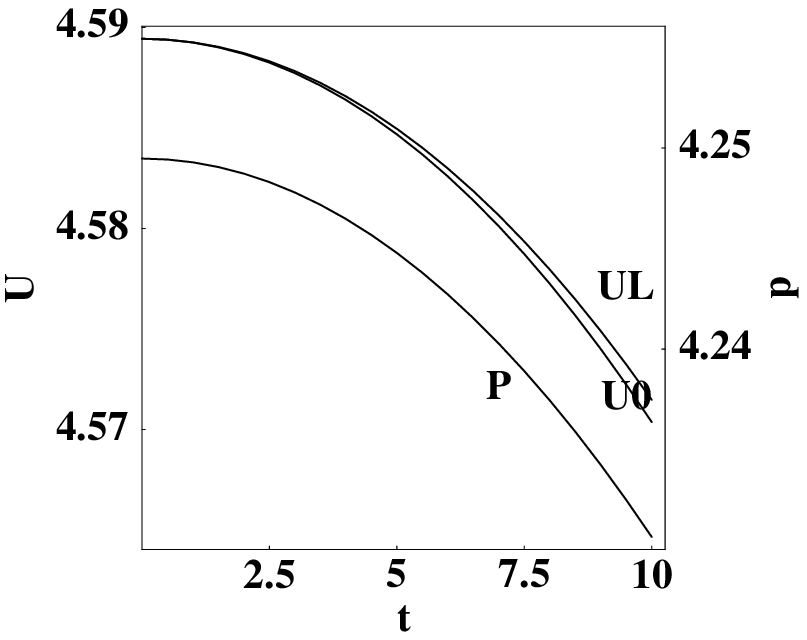}

\psfrag{d}[cc]{\rotatebox{180}{$10^{3}\,\delta/E_R$}}
\psfrag{t}[tc]{$\theta$ (deg)}
\psfrag{D}[br]{$\delta$}
\psfrag{K}[br]{$K$}
\psfrag{k}[bc]{\rotatebox{180}{$10^{3}\,K/E_R$}}

  \includegraphics[width=5.5cm]{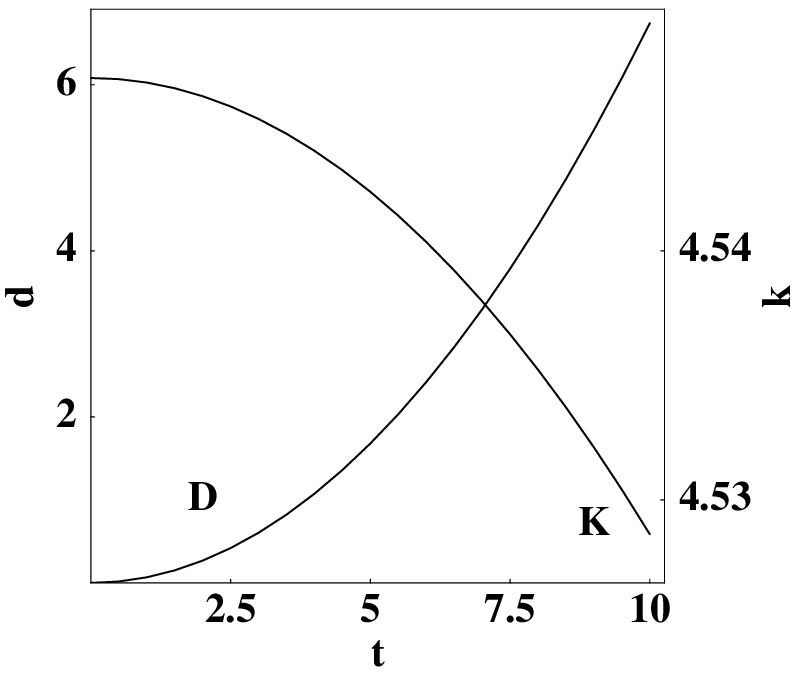}

\caption{Dependences of the parameters in the Bose-Hubbard
Hamiltonian (\ref{BHH}) on $\theta$
         for $^{87}$Rb, $q=\tilde\Omega_0^2/4\omega_R\Delta=-5$.}
\label{BHP}
\end{figure}

Together with the number density $ \hat{n} = \sum_{i=1}^L
(\hat{a}^\dagger_{0 i}\hat{a}_{0 i}+\hat{a}_{\Lambda
i}^\dagger\hat{a}_{\Lambda i})/L $, the $\hat {\bf T}$ generates
an U(2) algebra. However, for $g_a>0$ $(g_a<0)$ the Hamiltonian
(\ref{BHH}) only commutes with the generators $\hat{n}$ and
$\hat{T}_1$ ($\hat{T}_2$) of the respective U(1)$\times$U(1)
subgroups which they generate.

Increasing the angle $\theta$, one changes the form of dispersions
$E_\sigma(k)$ and Wannier spinors ${\bf W}_{\sigma i}(z)$, and
these changes are different for the $\Lambda$ and $0$ modes. As a
consequence, the values of the parameters in Hamiltonian
(\ref{BHH}) will be modified (see Ref.~\cite{Jaksch} for $
 J_0
 \left[
     \left(
     \tilde\Omega_0^2/\Delta
     \right)
     \cos\theta
 \right]
$, Ref.~\cite{KG03} for $J_\Lambda(\theta)$, and Fig.~\ref{BHP}
for the others) and relations (\ref{simpl}) become violated.
Therefore, varying the angle $\theta$ one can continuously tune
the parameters in the Hamiltonian (\ref{BHH}) and remove the
degeneracy by reducing the symmetry to the U(1) generated by
$\hat{n}$.

Throughout the paper in our numerical estimations we have used
known values of scattering lengths of $^{87}$Rb ($g_a<0$) and
$^{23}$Na ($g_a>0$)~\cite{Ho} and $\omega_\perp=2\pi\times 200$
Hz. The laser wavelengths $\lambda_L$ were chosen to be $780$ nm
and $600$ nm, respectively, which corresponds to the D1 line.

Despite the fact that Hamiltonian (\ref{BHH}) has been explicitly
derived here only for a one-dimensional lattice, it can be
realized in higher-dimensional optical lattices as well. This can
be done with the aid of additional pairs of linearly polarized
laser beams propagating along the $x$ and $y$ axes. In order to
avoid the interference of laser beams propagating in the
orthogonal directions their frequencies must be sufficiently
different. In order not to change the number of components in the
Bose-Hubbard Hamiltonian, the scheme of the internal atomic
transitions should remain the same. This can be achived if there
is no $\pi$ component of the laser field, i.e., all the laser
beams must be polarized in a plane perpendicular to the
quantization axis, which is chosen to be the $z$ axis. Therefore
the polarization vectors of the laser beams propagating along the
$y$ and $x$ axes are to be parallel to the $x$ and $y$ axes,
respectively. As discussed above and confirmed by numerical
calculations in the next section, the angle $\theta$ must be very
small and then $J_\Lambda \approx J_0$. Thus it is not difficult
to adjust the laser intensities and detunings of the additional
laser beams such that the anisotropy of the tunneling matrix
elements in the higher-dimensional version of the Hamiltonian
(\ref{BHH}) will be negligible.

\section{Mean-field approximation and the ground-state energy}

In order to investigate the superfluid--Mott-insulator phase
transition, we employ the mean-field
approximation~\cite{Sachdev,Oosten}
\begin{equation}
\label{mfa}
\hat a_{\sigma i}^\dagger
\hat a_{\sigma j}
\approx
\psi_\sigma
\left(
    \hat a_{\sigma j}
    +
    \hat a_{\sigma i}^\dagger
\right)
-
\psi_\sigma^2
\;,
\end{equation}
where $\psi_\sigma$ is the order parameter for Bose-Einstein
condensation in the component $\sigma$. The $\psi_\sigma$ carry
phases, which are in equilibrium either the same for the both
components or differ by $\pi$, as can be easily shown by energy
minimization. Thus we may take the $\psi_\sigma$ as real without
unwarranted loss of generality. In this approximation, the
Bose-Hubbard Hamiltonian becomes local and every lattice site is
described by the Hamiltonian
\begin{eqnarray}
\label{HBH1}
&&
\hat H_{BH}'
=
-
2 d
\sum_\sigma
    J_\sigma
\left[
    \left(
        \hat a_\sigma^\dagger
    +
    \hat a_\sigma
    \right)
    \psi_\sigma
    -
    \psi_\sigma^2
\right]
\\
&&
+
\sum_\sigma
\frac{U_\sigma}{2}
\hat n_\sigma (\hat n_\sigma -1)
+
K
\hat n_0 \hat n_\Lambda
\nonumber\\
&&
-
\frac{|P|}{2}
\left(
    \hat a_{0}^\dagger
    \hat a_{0}^\dagger
    \hat a_{\Lambda}
    \hat a_{\Lambda}
    +
    \hat a_{\Lambda}^\dagger
    \hat a_{\Lambda}^\dagger
    \hat a_{0}
    \hat a_{0}
\right)
-
\delta
\hat n_0
-
\mu
\sum_\sigma
\hat n_\sigma
\;,
\nonumber
\end{eqnarray}
where $d$ is the dimensionality of the lattice and we have omitted the site index $i$.
Since in the mean-field approximation $d$ plays a role of a numerical parameter,
we have done the calculations presented below for $d=1$. In the case of $d=2,3$,
there will be only numerical corrections, but all the qualitative
features will remain the same.

The values of the order parameters minimizing the ground-state energy
$E_g(\psi_0,\psi_\Lambda)$ of the Hamiltonian (\ref{HBH1}) determine
the phase diagram of the system at zero temperature:
$\psi_\sigma=0$ ($\psi_\sigma \neq 0$) corresponds to the
non-condensed (superfluid) phase of the $\sigma$-component.

\begin{figure}[t]
\centering

\psfrag{pl}[br]{$\psi_\Lambda$} \psfrag{p0}[br]{$\psi_0$}
\psfrag{e}[br]{$E_g/U$}
\psfrag{b}[c]{(a)} \psfrag{c}[c]{(b)}

  \includegraphics[width=4cm]{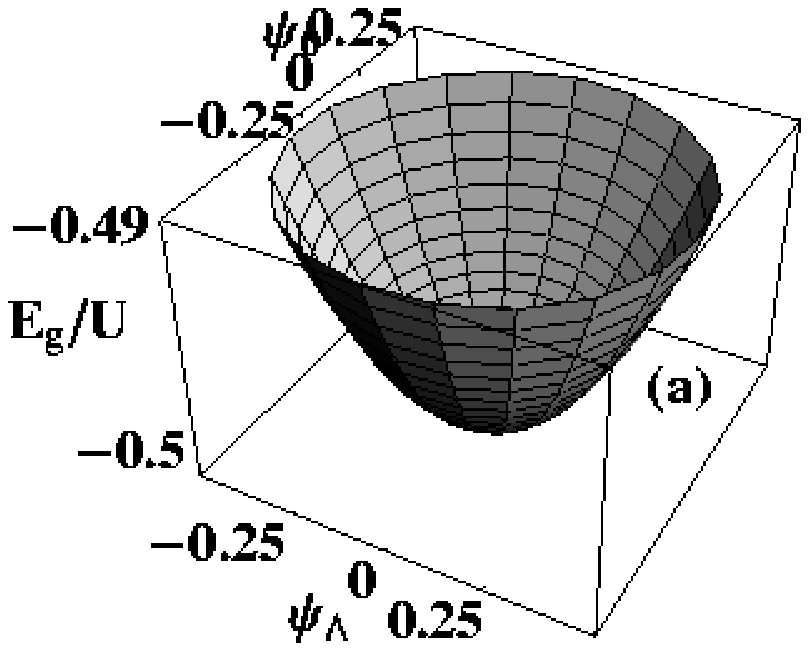} \hspace{2mm}
  \includegraphics[width=4cm]{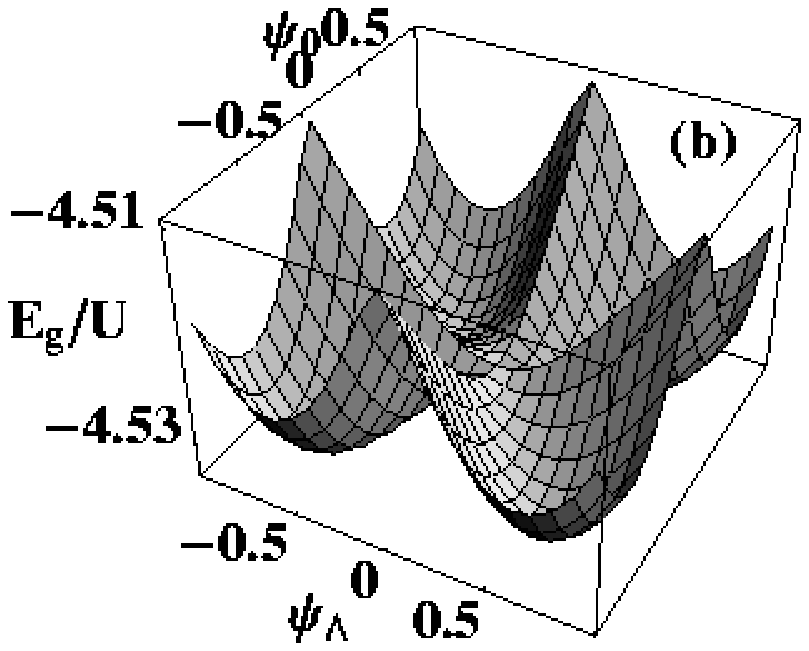}

\caption{Ground-state energy of the Hamiltonian (\ref{HBH1}) for
$^{87}$Rb ($g_a<0$)
         in the case $\theta=0$.
         $q=\tilde\Omega_0^2/4\omega_R\Delta=-4.5$,
     $\mu/U=0.5$ (a), $\mu/U=2.5$ (b).
        }
\label{eg0}
\end{figure}

The ground-state energy $E_g(\psi_0,\psi_\Lambda)$ has been
calculated by numerical diagonalization of the Hamiltonian
(\ref{HBH1}) using superpositions of the number states
\begin{equation}
|s\rangle
=
\sum_{n_0=0}^N
\sum_{n_\Lambda=0}^N
c_{n_0 n_\Lambda}
|n_0\rangle
|n_\Lambda\rangle
\;.
\end{equation}
We have done calculations with $N=10$ and the results are discussed below.

\subsection{Ferromagnetic interactions ($g_a<0$)}

\begin{figure}[b]
\centering

\psfrag{pl}[br]{$\psi_\Lambda$} \psfrag{p0}[br]{$\psi_0$}
\psfrag{e}[br]{$E_g/U_\Lambda$}

  \includegraphics[width=4cm]{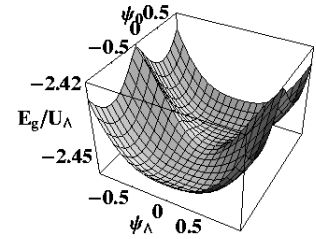}

\caption{Ground-state energy of the Hamiltonian (\ref{HBH1}) for
$^{87}$Rb ($g_a<0$)
         in the case $\theta=1^\circ$,
         $q=\tilde\Omega_0^2/4\omega_R\Delta=-4.5$,
     $\mu/U_\Lambda=1.71$.
        }
\label{eg1}
\end{figure}

\begin{figure*}[t]
\centering

\psfrag{p}[c]{\rotatebox{180}{$\psi_\Lambda$, $\psi_0$}}
\psfrag{g}[c]{$\theta=0$} \psfrag{h}[c]{$\theta=0.7^\circ$}
\psfrag{i}[c]{$\theta=1^\circ$}

  \includegraphics[width=5cm]{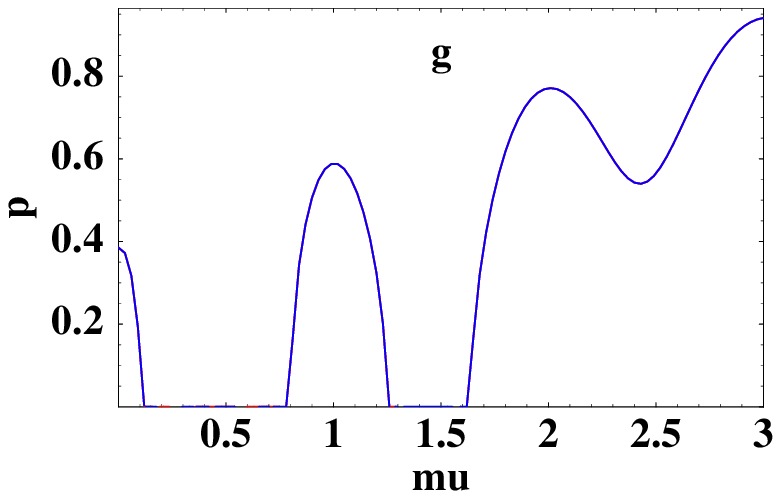}
  \includegraphics[width=5cm]{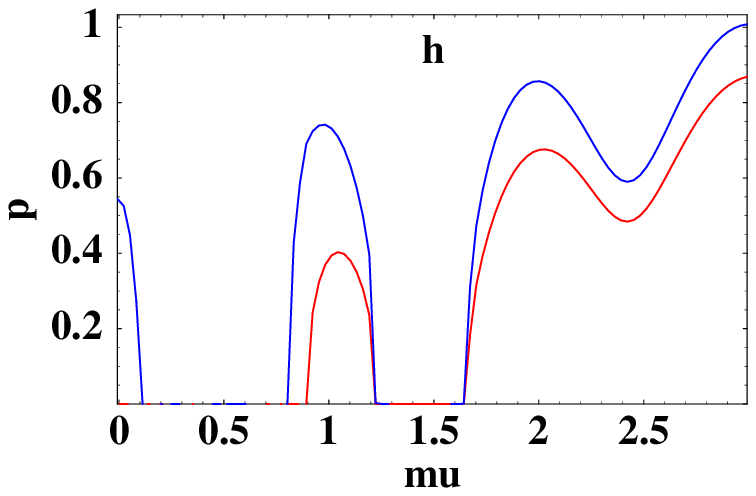}
  \includegraphics[width=5cm]{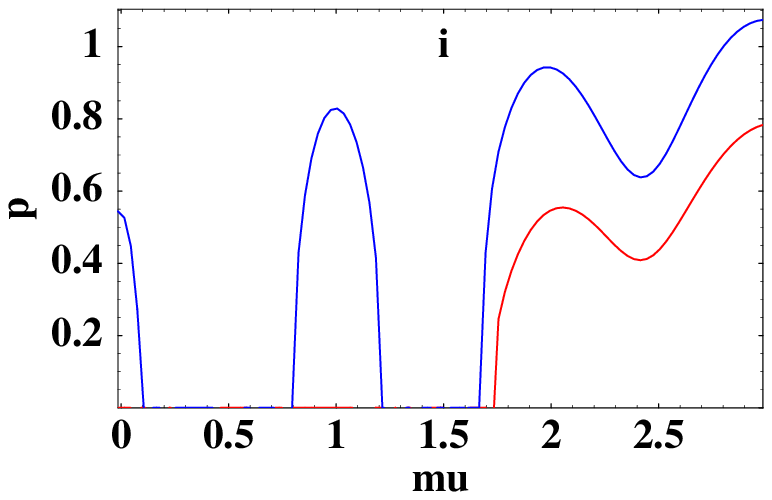}

\psfrag{mu}[t]{$\mu/U_\Lambda$}
\psfrag{n}[b]{\rotatebox{180}{$n_\Lambda$, $n_0$, $n$}}
\psfrag{t}[c]{$n$} \psfrag{=}[c]{$n_\Lambda=n_0$}
\psfrag{l}[c]{$n_\Lambda$} \psfrag{v}[c]{$n_0$}

  \includegraphics[width=5cm]{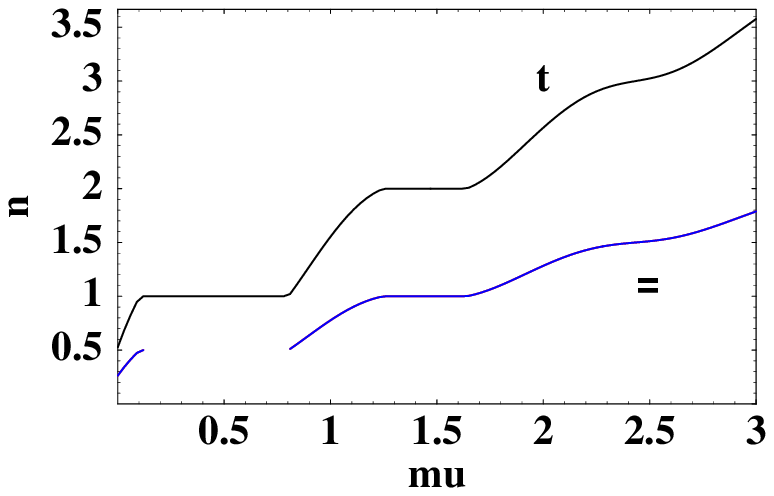}
  \includegraphics[width=5cm]{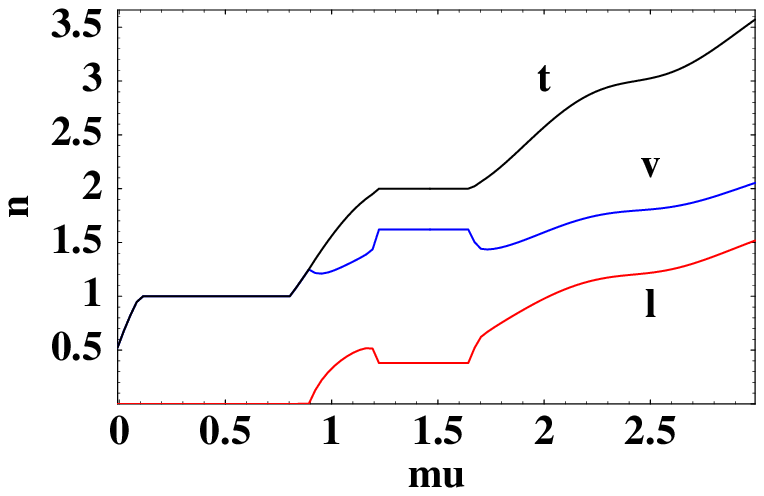}
  \includegraphics[width=5cm]{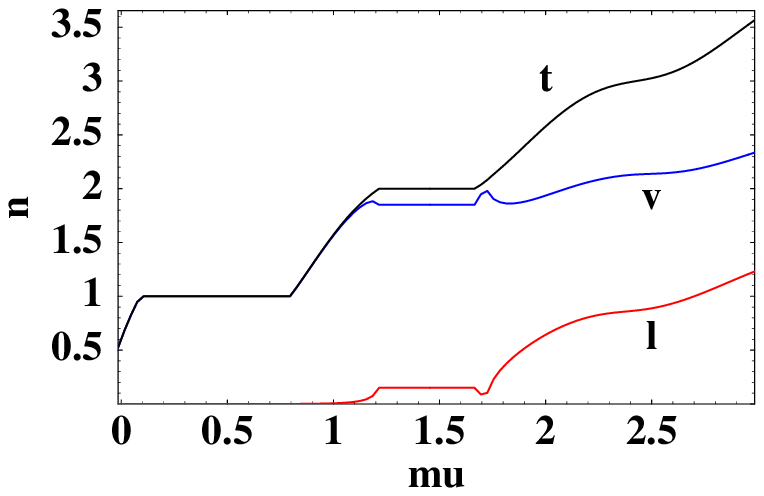}

\caption{Extremal values of the order parameters $\psi_\Lambda$,
$\psi_0$
         (upper row)
         and the mean occupation numbers of $\Lambda$ and $0$ components
     (lower row) for $^{87}$Rb ($g_a<0$).
         $q=-4.5$, $\theta=0$ (a), $0.7^\circ$ (b), $1^\circ$ (c)
        }
\label{psi}
\end{figure*}

\begin{figure}[b]
\centering

\psfrag{q}[c]{$q$}
\psfrag{mu}[c]{$\mu/U$}

  \includegraphics[width=5cm]{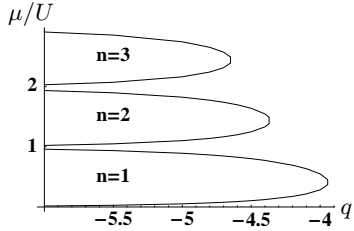}

\caption{Phase diagram for $^{87}$Rb ($g_a<0$) in the case $\theta=0$.
        }
\label{pdn}
\end{figure}

Typical dependences of the ground-state energy of the Hamiltonian
(\ref{HBH1}) on the order parameters $\psi_0$, $\psi_\Lambda$ for
$g_a<0$ are shown in Figs.~\ref{eg0}~and~\ref{eg1}. If $\theta=0$,
the ground-state energy $E_g$ is independent of the sign of the
real $\psi_\sigma$,
\begin{equation}
\label{sym} E_g(\psi_\Lambda,\psi_0)=E_g(|\psi_\Lambda|,|\psi_0|)
\;,
\end{equation}
and it is a symmetric function of $\psi_0$ and $\psi_\Lambda$:
$E_g(\psi_0,\psi_\Lambda)=E_g(\psi_\Lambda,\psi_0)$. More
generally, for complex $\psi_\sigma$ the isospin symmetry at
$\theta=0$ demands (still for $g_a<0$) that $E_g$ only depends on
the invariants $|\psi_0|^2+|\psi_\Lambda|^2$ and
$(\psi^*_\Lambda\psi_0+\psi^*_0\psi_\Lambda)$. One has to
distinguish the following two cases. (i) There is a single minimum
at $\psi_0=\psi_\Lambda=0$ [Mott phase, Fig.~\ref{eg0}(a),
corresponding to a state of unbroken U(1)$\times$U(1) symmetry].
(ii) There are four equal minima at $\psi_0^2=\psi_\Lambda^2\neq
0$ [symmetric two-component Bose-Einstein condensate (BEC),
Fig.~\ref{eg0}(b)]. In any of these minima the system has a
nonvanishing $x$ component of the expectation value of the
hyperfine spin density $ \langle{\bf f}_x\rangle \sim \langle
T_1\rangle \sim (\psi^*_\Lambda\psi_0+\psi^*_0\psi_\Lambda) /L $.

If $\theta \neq 0$ the scales of $\psi_0$ and $\psi_\Lambda$
become different. $E_g$ is not a symmetric function of $\psi_0$
and $\psi_\Lambda$ anymore, but Eq.~(\ref{sym}) remains valid. The
case (i) is possible again. Due to the different scalings of
$\psi_0$ and $\psi_\Lambda$ instead of (ii) we have $\psi_0^2 \neq
\psi_\Lambda^2\neq 0$ (two-component BEC). In addition, if $\theta
\neq 0$, there are additional cases with two equal minima of $E_g$
reached at (iii) $\psi_\Lambda=0$, $\psi_0 \neq 0$ [mixture of
noncondensed component $\Lambda$ and BEC of component $0$,
Fig.~\ref{eg1}]. As we see in Fig.~\ref{psi}, the mean occupation
numbers of the $\Lambda$ component turn out to be so small in
these regions that there seems to be no hope to observe the
noncondensed component experimentally.

The phase diagram in the case $\theta=0$ is shown in
Fig.~\ref{pdn}. The lobes correspond to the Mott phases with
different particle numbers per lattice site, and outside the lobes
we have a superfluid phase. The superfluid-insulator transition is
always second order for $g_a<0$.

\subsection{Antiferromagnetic interactions ($g_a>0$)}

\begin{figure}[b]
\centering

\psfrag{pl}[br]{$\psi_\Lambda$}
\psfrag{p0}[br]{$\psi_0$}
\psfrag{e}[br]{$E_g/U$}

\psfrag{a}[c]{(a)}
\psfrag{b}[c]{(b)}
\psfrag{c}[c]{(c)}

  \includegraphics[width=4cm]{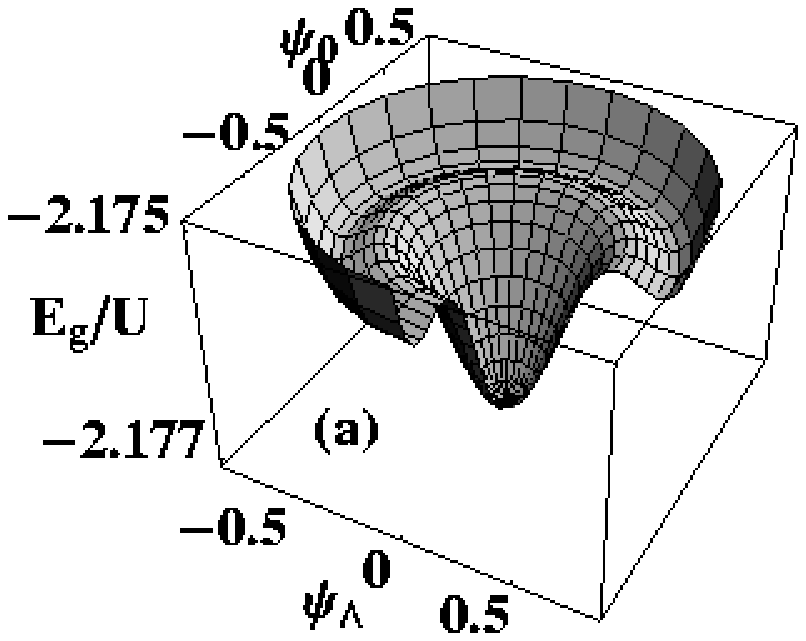}
  \includegraphics[width=4cm]{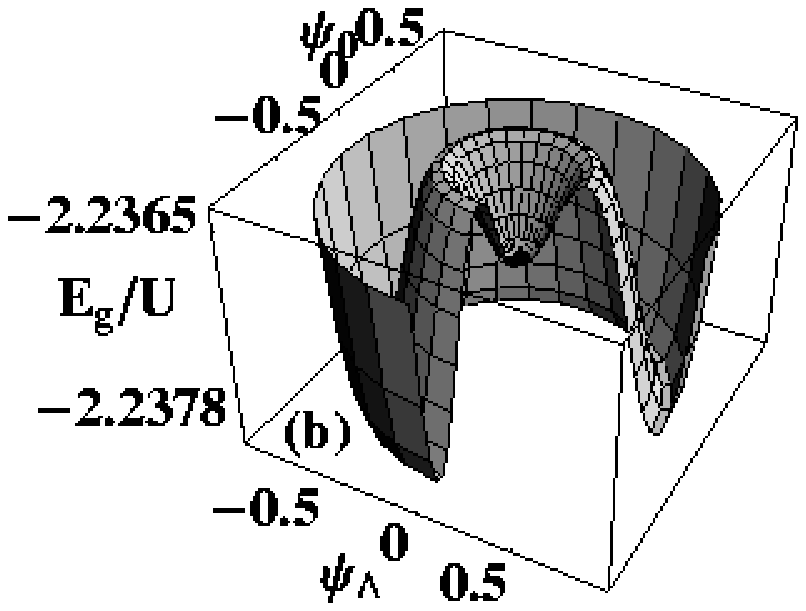}

  \includegraphics[width=4cm]{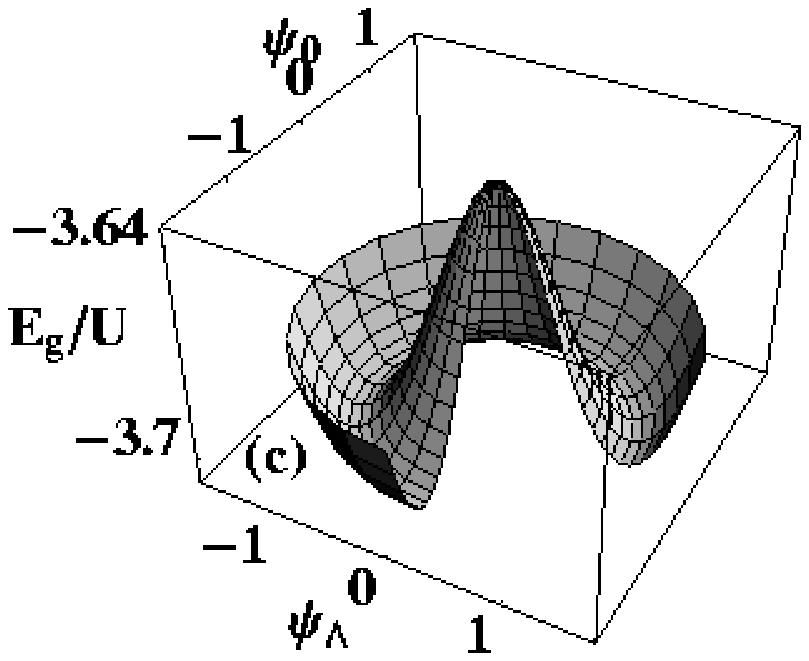}

\caption{Ground-state energy of the Hamiltonian (\ref{HBH1}) for
$^{23}$Na ($g_a>0$)
         in the case $\theta=0$.
         $q=\Omega_0^2/4\omega_R\Delta=-6.1$,
     $\mu/U=1.57$ (a), $\mu/U=1.6$ (b), $\mu/U=2.2$ (c).
        }
\label{eg0p}
\end{figure}

\begin{figure}[t]
\centering

\psfrag{pl}[br]{$\psi_\Lambda$}
\psfrag{p0}[br]{$\psi_0$}
\psfrag{e}[br]{$E_g/U$}

  \includegraphics[width=4cm]{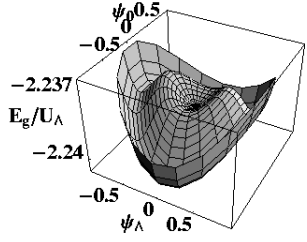}

\caption{Ground-state energy of the Hamiltonian (\ref{HBH1}) for
$^{23}$Na ($g_a>0$)
         in the case $\theta=0.1^\circ$.
         $q=\Omega_0^2/4\omega_R\Delta=-6.1$,
     $\mu/U_\Lambda=1.6$.
        }
\label{eg01p}
\end{figure}

\begin{figure}[b]
\centering

\psfrag{mu}[t]{}
\psfrag{p}[c]{\rotatebox{180}{$\psi_0$}}
\psfrag{t}[c]{$n$}
\psfrag{l}[c]{$n_\Lambda$}
\psfrag{v}[c]{$n_0$}

  \includegraphics[width=5cm]{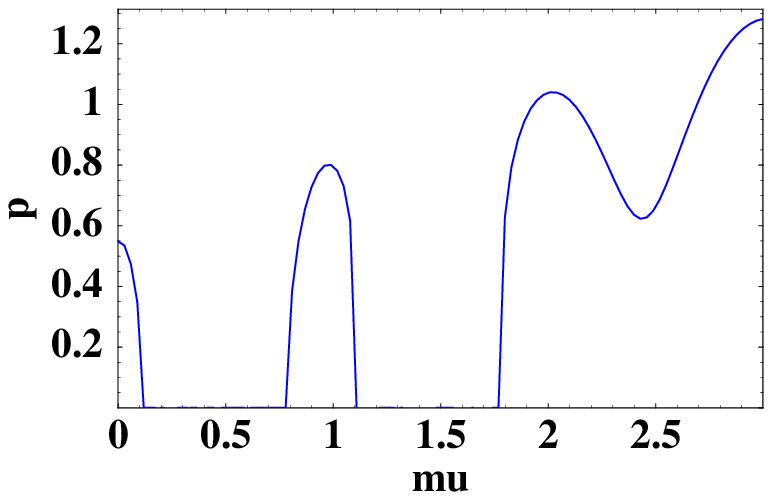}


\psfrag{mu}[t]{$\mu/U_\Lambda$}
\psfrag{n}[b]{\rotatebox{180}{$n_\Lambda$, $n_0$, $n$}}

  \includegraphics[width=5cm]{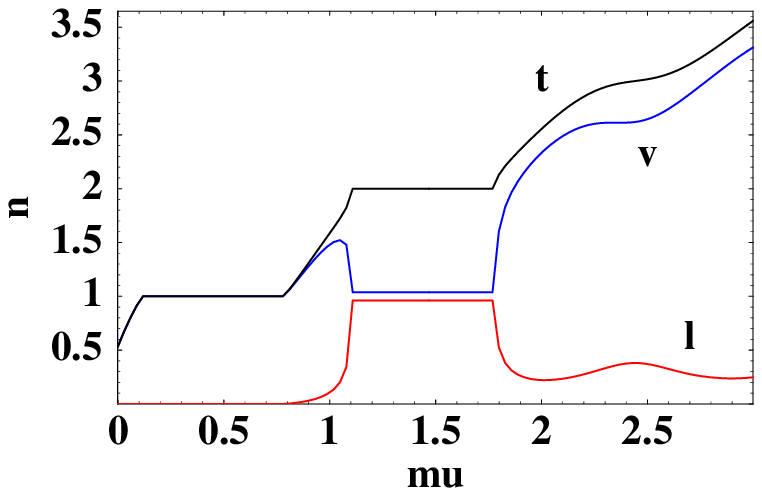}

\caption{Extremal values of the order parameter $\psi_0$
         and the mean occupation numbers of $\Lambda$ and $0$ components
     for $^{23}$Na ($g_a>0$).
         $q=-6.5$, $\theta=0.1^\circ$.
        }
\label{psip}
\end{figure}

In the case of positive $g_a$, i.e., positive $P$, the situation
is quite different. If $\theta=0$, the only condition in the
ground state is $\psi^*_0\psi_\Lambda-\psi^*_\Lambda\psi_0=0$,
which restricts the expectation value of the isospin $\langle{\bf
T}\rangle$ to the $(1,3)$ plane. Apart from this restriction $E_g$
is a rotationally symmetric function of the order parameters for
any values of $\mu$ and the dependences depicted in
Fig.~\ref{eg0}(b) do not occur. Case (i) of the previous
subsection remains possible, but instead of (ii) the minima of
$E_g$ are located on a circle $\psi_0^2+\psi_\Lambda^2=\psi^2$ on
which a point is selected by spontaneous symmetry breaking [stable
two-component Bose-Einstein condensed phase with spontaneously
broken isospin symmetry with $\langle{\bf T}\rangle$ oriented
arbitrarily in the $(1,3)$ plane, Fig.~\ref{eg0p}(c)]. In
addition, there can be not only global minima, but also local ones
either at $\psi_\Lambda=\psi_0=0$ [metastable Mott phase,
Fig.~\ref{eg0p}(b)] or on a circle [metastable superfluid phase,
Fig.~\ref{eg0p}(a)], depending on the system parameters. This
means that in some parameter regions the superfluid and Mott
phases can coexist and the phase transition can be either first or
second order in contrast to the case of ferromagnetic
interactions, where it is always second order.

If $\theta \neq 0$, the  symmetry between the two components is destroyed and the scalings
for $\psi_0$ and $\psi_\Lambda$ become different. The minima of $E_g$ can be
reached at either $\psi_0=\psi_\Lambda=0$ or $\psi_\Lambda=0$, $\psi_0 \neq 0$
as in the case $g_a<0$ discussed above. In addition, there are dependences
shown in Fig.~\ref{eg01p} with (or without) one local minimum at $\psi_\Lambda=\psi_0=0$ and
two global minima at $\psi_\Lambda=0$, $\psi_0 \neq 0$.
In contrast to the case of negative $g_a$,
the populations of the components can be comparable in all the cases
(Fig.~\ref{psip}).

\begin{figure}[t]
\centering

\psfrag{q}[c]{$q$}
\psfrag{mu}[c]{$\mu/U$}

  \includegraphics[width=5cm]{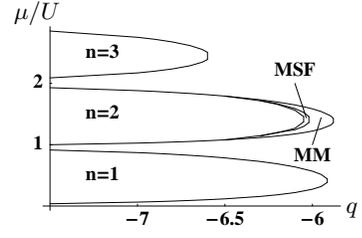}

\caption{Phase diagram for $^{23}$Na ($g_a>0$) in the case $\theta=0$.
         The regions of metastable superfluid phase coexisting with the stable
     Mott phase and that of matastable Mott phase coexisting with
     the stable superfluid phase are denoted by MSF and MM, respectively.
        }
\label{pdp}
\end{figure}

The phase diagram in the case $\theta=0$ is shown in
Fig.~\ref{pdp}. It consists of a series of lobes corresponding to
the Mott phase as in the case of ferromagnetic interactions.
However, the superfluid-insulator transition for $n=2$ is now
first order, while for $n=1,3$ it remains to be second order. In
the case $n=2$, the stable (metastable) superfluid phase coexists
with metastable (stable) Mott phase in a small region near the
phase boundary.

\subsection{Mott phase, arbitrary $g_a$}

In the Mott phase, $\psi_\Lambda=\psi_0=0$ and for every number of
atoms per site $n$ there are $n+1$ eigenstates of the local
Hamiltonian (\ref{HBH1}). For $n \le 3$ the expressions for
eigenstates are very simple. If the number of particles per site
$n$ is less than two, the terms which contain $|P|$ in
Eq.~(\ref{HBH1}) do not give any contribution and we have the
usual Fock states,
\begin{eqnarray}
E_0
&=&
0
\;,
\quad
|n=0\rangle = |0\rangle_\Lambda |0\rangle_0
\;,
\nonumber\\
E_{1,0}
&=&
-(\mu+\delta)
\;,
\quad
|n=1\rangle_0
=
|0\rangle_\Lambda |1\rangle_0
\;,
\nonumber\\
E_{1,1}
&=&
-\mu
\;,
\quad
|n=1\rangle_1
=
|1\rangle_\Lambda
|0\rangle_0
\;.
\end{eqnarray}
The lowest of these energies gives the ground-state energy per
lattice site. The states with higher energies correspond to
excited states whose positive excitation energies are the
difference of the respective energies $E$ and that of the ground
state. In the case $\theta=0$, $\delta$ vanishes. The states
$|n=1\rangle_0$ and $|n=1\rangle_1$ become degenerate and any
linear combination of these states forms the SU(2)-symmetric
manifold of degenerate ground states in the case of one atom per
lattice site. This case requires a positive chemical potential
$\mu\ge 0$, and, as we shall see by comparing with the energy
levels for the case of two atoms per lattice site, $\mu\le U-|P|$.
Thus, in the Mott phase with one atom per site, the mean numbers
of atoms in $\Lambda$ and $0$ modes $n_\Lambda$ and $n_0$ are
undefined, but the total number of atoms $n=n_\Lambda+n_0$ is
fixed [see Fig.~\ref{psi}(a)]. If $\theta \neq 0$, the degeneracy
is removed and $n_\Lambda=0$, $n_0=1$
[Figs.~\ref{psi}(b)~and~(c)].

If $n \ge 2$, the eigenstates are in general superpositions of Fock states.
For $n=2$ we get
\begin{eqnarray}
&&
E_{2,1}
=
K-\delta-2\mu
\;,
\quad |n=2\rangle_1
=
|1\rangle_\Lambda |1\rangle_0
\\
&&
E_{2,\pm}
=
\frac{U_0+U_\Lambda}{2}
-
\delta
\pm
\sqrt{
       \left(
           \frac{U_0-U_\Lambda}{2}
       -\delta
       \right)^2
       +
       P^2
     }
-2\mu
,
\nonumber\\
&&
|n=2\rangle_\pm = c_{0\pm}^{(2)}
\;
|0\rangle_\Lambda
|2\rangle_0
+
c_{2\pm}^{(2)}
\;
|2\rangle_\Lambda |0\rangle_0
\nonumber\\
&&
c_{0,\pm}^{(2)}
=
\left\{
    \frac{1}{2}
    \left[
        1
    \pm
    \frac
    {U_0-U_\Lambda-2\delta}
    {
     \sqrt{
                \left(
                    U_0-U_\Lambda-2\delta
        \right)^2
        +
        4 P^2
          }
    }
    \right]
\right\}^{1/2}
\quad,\quad
\nonumber\\
&&
c_{2,\pm}^{(2)}
=
\mp
\left\{
    \frac{1}{2}
    \left[
        1
    \mp
    \frac
    {U_0-U_\Lambda-2\delta}
    {
     \sqrt{
                \left(
                    U_0-U_\Lambda-2\delta
        \right)^2
        +
        4 P^2
          }
    }
    \right]
\right\}^{1/2}
\;.
\nonumber
\end{eqnarray}
In the case $\theta=0$, $c_{0,\pm}^{(2)}=1/\sqrt{2}$ and
$c_{2,\pm}^{(2)}=\mp 1/\sqrt{2}$. For $\mu\ge U-|P|$ the ground
state is either $|n=2\rangle_-$ or an arbitrary linear combination
of $|n=2\rangle_1$ and $|n=2\rangle_-$: If $g_a>0$,
$|n=2\rangle_-$ is a ground state and the mean particle numbers
$n_\Lambda=n_0=1$. If $g_a<0$,
$E_{2,-}=E_{2,1}=U-|P|-2\mu<E_{2,+}$, i.e., the ground state is
doubly degenerate. Nevertheless, we have again $n_\Lambda=n_0=1$.

For $n=3$
\begin{eqnarray}
&&
E_{3g\pm}
=
\frac{3U_0+U_\Lambda}{2} + K - 2\delta
\nonumber\\
&&
\pm
\sqrt{
       \left(
           \frac{3 U_0-U_\Lambda}{2}
       - K - \delta
       \right)^2
       +
       3 P^2
     }
-3\mu
\nonumber\\
&&
|n=3\rangle_{g\pm}
=
c_{0\pm}^{(3)} \; |0\rangle_\Lambda
|3\rangle_0
+
c_{2\pm}^{(3)} \; |2\rangle_\Lambda |1\rangle_0 \;,
\nonumber\\
&&
\hspace{-5mm}
c_{0,\pm}^{(3)}
=
\left\{
    \frac{1}{2}
    \left[
        1
    \pm
    \frac
    {3 U_0-U_\Lambda-2K-2\delta}
    {
     \sqrt{
                \left(
                    3U_0-U_\Lambda-2K-2\delta
        \right)^2
        +
        12 P^2
          }
    }
    \right]
\right\}^{1/2}
\\
&&
\hspace{-5mm}
c_{2,\pm}^{(3)}
=
\mp
\left\{
    \frac{1}{2}
    \left[
        1
    \mp
    \frac
    {3 U_0-U_\Lambda-2K-2\delta}
    {
     \sqrt{
                \left(
                    3U_0-U_\Lambda-2K-2\delta
        \right)^2
        +
        12 P^2
          }
    }
    \right]
\right\}^{1/2}
\nonumber\\
&&
E_{3u\pm}
=
\frac{U_0+3U_\Lambda}{2}
+ K - \delta
\nonumber\\
&&
\pm
\sqrt{
       \left(
           \frac{U_0-3U_\Lambda}{2}
       + K - \delta
       \right)^2
       +
       3 P^2
     }
-3\mu
\nonumber\\
&&
|n=3\rangle_{u\pm}
=
c_{1\pm}^{(3)}
\;
|1\rangle_\Lambda
|2\rangle_0
+
c_{3\pm}^{(3)}
\;
|3\rangle_\Lambda
|0\rangle_0
\;,
\nonumber\\
&&
\hspace{-5mm}
c_{1,\pm}^{(3)}
=
\left\{
    \frac{1}{2}
    \left[
        1
    \pm
    \frac
    {U_0-3U_\Lambda+2K-2\delta}
    {
     \sqrt{
                \left(
                    U_0-3U_\Lambda+2K-2\delta
        \right)^2
        +
        12 P^2
          }
    }
    \right]
\right\}^{1/2}
\nonumber\\
&&
\hspace{-5mm}
c_{3,\pm}^{(3)}
=
\mp
\left\{
    \frac{1}{2}
    \left[
        1
    \mp
    \frac
    {U_0-3U_\Lambda+2K-2\delta}
    {
     \sqrt{
                \left(
                    U_0-3U_\Lambda+2K-2\delta
        \right)^2
        +
        12 P^2
          }
    }
    \right]
\right\}^{1/2}
\nonumber
\end{eqnarray}
In the case $\theta=0$, the states are doubly degenerate and we
have $E_{3g-}=E_{3u-}<E_{3g+}=E_{3u+}$. If $g_a<0$, the
coefficients are $c_{0-}^{(3)}=c_{3-}^{(3)}=1/2$,
$c_{1-}^{(3)}=c_{2-}^{(3)}=\sqrt{3}/2$, and in spite of the
degeneracy of the ground state we have $n_\Lambda=n_0=3/2$. If
$g_a>0$, the coefficients are
$c_{0-}^{(3)}=c_{3-}^{(3)}=\sqrt{3}/2$,
$c_{1-}^{(3)}=c_{2-}^{(3)}=1/2$, and the mean particle numbers
$n_\Lambda$, $n_0$ become undetermined. The degenerate ground
states $|n=3\rangle_{g-}$ and $|n=3\rangle_{u-}$ occur in a domain
for $\mu\ge 2(U-|P|)$, if $g_a\le 0$, and for $\mu\ge 2U$, if
$g_a\ge 0$.

\section{Collective excitations in the superfluid phase}

From the Hamiltonian~(\ref{BHH}) we obtain the following Heisenberg
equations of motion:

\begin{eqnarray}
\label{he}
i\hbar
\frac{d}{dt}
\hat a_{\Lambda i}
&=&
-J_\Lambda
\left(
    \hat a_{\Lambda, i-1}
    +
    \hat a_{\Lambda, i+1}
\right)
+
U_\Lambda
\hat n_{\Lambda i}
\hat a_{\Lambda i}
\nonumber\\
&+&
K
\hat n_{0 i}
\hat a_{\Lambda i}
-
\left|
    P
\right|
\hat a_{\Lambda i}^\dagger
\hat a_{0 i}
\hat a_{0 i}-\mu\hat{a}_{\Lambda i}
\;,
\\
i\hbar
\frac{d}{dt} \hat a_{0 i}
&=&
-J_0
\left(
    \hat a_{0, i-1}
    +
    \hat a_{0, i+1}
\right)
+
U_0
\hat n_{0 i}
\hat a_{0 i}
\nonumber\\
&+&
K
\hat n_{\Lambda i}
\hat a_{0 i}
-
\left|
    P
\right|
\hat a_{0 i}^\dagger
\hat a_{\Lambda i}
\hat a_{\Lambda i}
-
(\delta+\mu)
\hat a_{0 i}
\;.
\nonumber
\end{eqnarray}
These equations remain valid in higher-dimensional lattices as well with the only
modification that the lattice indices aquire a vector character.

In the mean-field approximation, the operators $\hat a_{\sigma i}$
are replaced by position independent c numbers $\alpha_\sigma$:
$
 \hat a_{\sigma i}
 \approx
 \alpha_\sigma
$.
One then finds a mean-field equation of the classical Hamiltonian form
\begin{equation}
i\hbar\frac{d}{dt}\alpha_\sigma
=
\frac
{\partial H(\alpha_0,\alpha_\Lambda,\alpha_0^*,\alpha_\Lambda^*)}
{\partial \alpha_\sigma^*}
\end{equation}
with the local classical mean-field Hamiltonian
\begin{eqnarray}
&&
H
=
-
2
\left(
    J_0
    |\alpha_0|^2
    +
    J_\Lambda|\alpha_\Lambda|^2
\right)
+
\frac{U_0}{2}
|\alpha_0|^4
+
\frac{U_\Lambda}{2}
|\alpha_\Lambda|^4
\nonumber\\
&&
+
K
|\alpha_0\alpha_\Lambda|^2
-
\frac{|P|}{2}
\left[
    (\alpha_\Lambda^*\alpha_0)^2
    +
    (\alpha_0^*\alpha_\Lambda)^2
\right]
\nonumber\\
&&
-
\left(
    \delta
    +
    \mu
\right)
|\alpha_0|^2
-
\mu
|\alpha_\Lambda|^2
\;.
\end{eqnarray}
At least roughly, the $\alpha_\sigma$ correspond to the previously
introduced order parameters $\psi_\sigma$, but it should be noted
that the present mean-field approximation is different from the
one of the previous section [Eq.(\ref{mfa})], where it was
employed only to approximate the hopping term of the Hamiltonian,
while now we use it equally in all terms of the Hamiltonian. The
difference of these approximations directly corresponds to the
difference of our goals in this and the preceeding section: there
we aimed at a description of the Mott state, where the
$\psi_\sigma$ both vanish, and its eventual instabilty toward the
formation of a superfluid state with an incipient (small)
nonvanishing $\psi_\sigma$, while here our goal is to describe the
excitations deep in the superfluid states, where the mean field is
large and contains practically all particles. The price we have to
pay for achieving the latter goal is that the Mott state cannot be
described  by the modified form of the mean-field approximation.

For $\theta=0$ we can use Eqs.~(\ref{simpl}) to simplify
the local mean-field Hamiltonian, which can then be written as
\begin{equation}
H
=
-
(\mu+2J)
n
+
\frac{U}{2}n^2
-
\frac{|P|}{2}
\left(
    \alpha_\Lambda^*
    \alpha_{0}\pm
    \alpha_0^*
    \alpha_\Lambda
\right)^2
\;,
\end{equation}
where ``$+$" and ``$-$" correspond to $g_a<0$ and $g_a>0$, respectively.

In the equlibrium state, we set the time derivatives of the mean
fields to zero. Then it is easy to verify that the phases of
$\alpha_0$ and $\alpha_\Lambda$ must be equal, and may be set to
zero for convenience, i.e., we may, for the present purpose,
simplify the mean-field ansatz
\begin{equation}
\label{amf}
\hat a_{\sigma i}
\approx
\sqrt{n_\sigma}
\;,
\end{equation}
with the normalization condition $n_\Lambda+n_0=n$.
It is the small but nonzero value of $|P|$ which aligns the phases
of the $\hat{a}_{\sigma i}$ for $\sigma=0, \Lambda$. We are then
left with the equations
\begin{eqnarray}
\label{mune}
\mu
\sqrt{n_\Lambda}
&=&
\left(
    - 2 J_\Lambda + U_\Lambda n_\Lambda + K n_0 - |P| n_0
\right)
\sqrt{n_\Lambda}
\;,
\\
\mu
\sqrt{n_0}
&=&
\left(
    - 2 J_0 + U_0 n_0 + K n_\Lambda - |P| n_\Lambda - \delta
\right)
\sqrt{n_0}
\;.
\nonumber
\end{eqnarray}

We consider the case $g_a>0$ first. Then eqs.(\ref{mune}) in
general have no positive solutions for $n_0$ and $n_\Lambda$ with
$n=n_0+n_\Lambda$. However, there is an exception for the special
case $\theta=0$ where, as we have discussed, the usual U(1)
symmetry of particle number conservation is enhanced to a
U(1)$\times$U(1) symmetry. As it was also mentioned above, it is
in any case reasonable to consider only small values of $\theta$.
Since the variation of $\delta$ with $\theta$ is much faster than
that of $U_0$, $U_\Lambda$, and $P$, for the semiquantitative
analysis one can put $U_\Lambda \approx U_0$ and $K \approx U_0 +
P$. Then Eqs.~(\ref{mune}) take the form
\begin{eqnarray}
\label{mune1}
\mu
\sqrt{n_\Lambda}
&=&
\left(
    - 2 J_\Lambda + U_0 n
\right)
\sqrt{n_\Lambda}
\;,
\\
\label{mune2}
\mu
\sqrt{n_0}
&=&
\left(
    - 2 J_0 + U_0 n - \delta
\right)
\sqrt{n_0}
\;.
\end{eqnarray}
Let us assume that $n_\Lambda \ne 0$, $n_0 \ne 0$. Then we can
divide Eqs.~(\ref{mune1}) and (\ref{mune2}) by $\sqrt{n_0}$,
$\sqrt{n_\Lambda}$, respectively, and we see immediately that
Eqs.~(\ref{mune1}),(\ref{mune2}) are compartible only if
$J_\Lambda=J_0+\delta/2$, which does occur for $\theta=0$.
Therefore in the degenerate case $\theta=0$ only the total density
$n=n_0+n_\Lambda$ is determined,
\begin{equation}
\mu = U n - 2 J
\;,
\end{equation}
and the actual values taken by $n_0$ and $n_\Lambda$ break
spontaneously the SU(2) symmetry by fixing an arbitrary
orientation of ${\bf T}$ in the $(1,3)$ plane. For $\theta \ne 0$,
either $n_\Lambda$ or $n_0$ must vanish, i.e., ${\bf T}$ is then
oriented parallel or antiparallel to the 3 direction,
respectively. Since the $0$ mode has a lower energy, the solution
of Eqs.~(\ref{mune1})~and~(\ref{mune2}) is
\begin{equation}
\label{n0} n_0=n
\;,
\quad
n_\Lambda=0
\;,
\quad
\mu + \delta
=
U_0 n - 2 J_0
\;,
\end{equation}
exactly as in the case of one-component BEC~\cite{Oosten,Rey}.

In the case $g_a<0$, the situation is different and the solution
of Eq.~(\ref{mune}) has the form
\begin{widetext}
\begin{eqnarray}
\label{nm}
n_\Lambda
&=&
\frac
{
  \left(
      U_0 - K + |P|
  \right)
  n
  -
  2
  \left(
      J_0
      -
      J_\Lambda
  \right)
  -
  \delta
}
{
  U_\Lambda + U_0
  -
  2
  \left(
      K-|P|
  \right)
}
\;,\quad
n_0
=
\frac
{
  \left(
      U_\Lambda - K + |P|
  \right)
  n
  +
  2
  \left(
      J_0
      -
      J_\Lambda
  \right)
  +
  \delta
}
{
  U_\Lambda + U_0
  -
  2
  \left(
      K-|P|
  \right)
}
\;,
\\
\mu
&=&
-
\left(
    J_0
    +
    J_\Lambda
\right)
+
\frac
{
  2
  \left(
      J_0
      -
      J_\Lambda
  \right)
  \left(
      U_0
      -
      U_\Lambda
  \right)
  -
  \delta
  \left(
      U_\Lambda + |P| - K
  \right)
  +
  \left[
      U_0 U_\Lambda
      -
      \left(
          K - |P|
      \right)^2
  \right]
  n
}
{
  U_\Lambda + U_0
  -
  2
  \left(
      K-|P|
  \right)
}
\;,
\nonumber
\end{eqnarray}
\end{widetext}
provided that
\begin{equation}
\label{ineq}
\left(
    U_0 - K + |P|
\right)
n
-
2
\left(
    J_0
    -
    J_\Lambda
\right)
-
\delta
>
0
\;,
\end{equation}
otherwise the solution is given by Eq.~(\ref{n0}). For $n_0$ and
$n_\Lambda$ both nonvanishing the Bose gas has a nonvanishing
hyperfine and magnetization density in the $x$ direction
proportional to $\sqrt{n_0n_\Lambda}$.

In the case $g_a<0$, $\theta=0$, Eqs.~(\ref{nm}) take a very
simple form,
\begin{equation}
n_0=n_\Lambda=\frac{n}{2}
\;,
\quad
\mu
=
\left(
    U - |P|
\right)
n
-
2 J
\;.
\end{equation}
The excitations can be worked out employing the ansatz~\cite{Oosten,Rey}
\begin{equation}
\label{ansatz}
\hat a_{\sigma i}
=
    \sqrt{n_\sigma}
    +
    \hat \Delta_{\sigma i}
\;.
\end{equation}
Substituting Eq.~(\ref{ansatz}) into Eq.~(\ref{he}) we get the
following linearized equations for the excitations:
\begin{eqnarray}
\label{Delta}
&&
i\hbar
\frac{d}{dt}\hat \Delta_{\Lambda i}
=
-J_\Lambda
\left(
    \hat \Delta_{\Lambda, i-1}
    +
    \hat \Delta_{\Lambda, i+1}
\right)
\nonumber\\
&&+
\left(
    2 U_\Lambda n_\Lambda
    +
    K n_0
    -
    \mu
\right)
\hat \Delta_{\Lambda i}
+
\left(
    U_\Lambda n_\Lambda
    -
    |P| n_0
\right)
\hat \Delta_{\Lambda i}^\dagger
\nonumber\\
&&+
\left(
    K - 2|P|
\right)
\sqrt{n_0 n_\Lambda}
\hat \Delta_{0 i}
+
K
\sqrt{n_0 n_\Lambda}
\hat \Delta_{0 i}^\dagger
\;,
\\
&&
i\hbar
\frac{d}{dt} \hat \Delta_{0 i}
=
-J_\Lambda
\left(
    \hat \Delta_{0, i-1}
    +
    \hat \Delta_{0, i+1}
\right)
\nonumber\\
&&+
\left(
    2 U_0 n_0
    +
    K n_\Lambda
    -
    \mu
    -
    \delta
\right)
\hat \Delta_{0 i}
+
\left(
    U_0 n_0
    -
    |P| n_\Lambda
\right)
\hat \Delta_{0 i}^\dagger
\nonumber\\
&&+
\left(
    K - 2|P|
\right)
\sqrt{n_0 n_\Lambda}
\hat \Delta_{\Lambda i}
+
K
\sqrt{n_0 n_\Lambda}
\hat \Delta_{\Lambda i}^\dagger
\;.
\nonumber
\end{eqnarray}
From Eqs.~(\ref{Delta}) we obtain two coupled wave equations for the density
fluctuations
$
  \hat\rho_{\sigma i}
  =
  \sqrt{n_\sigma}
  \left(
      \hat \Delta_{\sigma i}
      +
      \hat \Delta_{\sigma i}^\dagger
  \right)
$,
\begin{eqnarray}
\label{qwe}
-
\hbar^2
\frac{d^2}{dt^2}
\hat\rho_{\sigma i}
&=&
J_\sigma^2
\left(
    \hat\rho_{\sigma,i-2}
    +
    \hat\rho_{\sigma,i+2}
\right)
\\
&+&
A_{\sigma\sigma}^{(1)}
\left(
    \hat\rho_{\sigma,i-1}
    +
    \hat\rho_{\sigma,i+1}
\right)
+
A_{\sigma\sigma}^{(0)}
\hat\rho_{\sigma i}
\nonumber\\
&+&
A_{\sigma\sigma'}^{(1)}
\left(
    \hat\rho_{\sigma',i-1}
    +
    \hat\rho_{\sigma',i+1}
\right)
+
A_{\sigma\sigma'}^{(0)}
\hat\rho_{\sigma' i}
\nonumber
\;.
\end{eqnarray}
The coefficients $A_{\sigma \sigma'}^{(0,1)}$ are listed in the
Appendix. The eigenmodes can be found by means of Fourier
transformation,
\begin{equation}
\hat\rho_{\sigma i}
=
\sum_k
c_{\sigma k}
\hat b_k
\exp
\left[
    i
    \left(
        k z_i - \omega_k t
    \right)
\right]
\;.
\end{equation}
In the case $\theta=0$, the two wave equations for the sum and the difference of
the two densities are decoupled. The coefficients $c_{\sigma k}$ are then independent
of $k$. For $g_a<0$ the two eigenmodes are given by
\begin{eqnarray}
\hbar\omega_k^{(+)}
&=&
\sqrt
{
  {\cal E}(k)
  \left[
      {\cal E}(k)
      +
      2
      \left(
          U - |P|
      \right)
      n
  \right]
}
\;,\;
c_\Lambda^{(+)}/c_0^{(+)}=1
\;,
\nonumber\\
\hbar\omega_k^{(-)}
&=&
{\cal E}(k)
+
2|P|n
\;,\qquad
c_\Lambda^{(-)}/c_0^{(-)}=-1
\;,
\end{eqnarray}
and for $g_a>0$ by
\begin{eqnarray}
\label{Gold}
\hbar\omega_k^{(+)}
&=&
\sqrt
{
  {\cal E}(k)
  \left[
      {\cal E}(k)
      +
      2 U n
  \right]
}
\;,\quad
c_\Lambda^{(+)}/c_0^{(+)}=1
\;,
\\
\hbar\omega_k^{(-)}
&=&
\sqrt
{
  {\cal E}(k)
  \left[
      {\cal E}(k)
      +
      2 P n
  \right]
}
\;,\quad
c_\Lambda^{(-)}/c_0^{(-)}=-1
\;,
\nonumber
\end{eqnarray}
where
\begin{equation}
\label{ld}
{\cal E}(k)=4J\sin^2
\left(
    \frac{\pi}{2}\frac{k}{k_L}
\right)
\end{equation}
is a dispersion relation of free particles in a lattice. The
symmetric modes with the eigenfrequencies $\omega_k^{(+)}$ are
qualitatively the same in both cases ($g_a<0$ and $g_a>0$). They
are the Goldstone modes of the spontaneous breaking of the U(1)
symmetry of particle number conservation, which correspond to
condensate excitations. The antisymmetric mode is principally
different: In the case $g_a>0$, it is an additional Goldstone mode
[in which the components $\delta T_\perp$ transverse to the
isospin vector in the $(1,3)$ plane oscillate] corresponding to
the spontaneous breaking of the isospin symmetry, by the ground
state; but for $g_a<0$ the isospin is oriented along the
1-direction by the interaction and the isospin symmetry around
this axis is not spontaneously broken. In the antisymmetric mode
there is therefore a gap of size $2|P|n$, and
$(\omega_k^{(-)}-2|P|n) \sim k^2$ for small $k$. This is again an
isospin wave with $\delta T_2$ and $\delta T_3$ propagating
together. In the case $g_a<0$, the gap allows the energies of the
symmetric and antisymmetric modes to cross at the points
\begin{equation}
\label{cp}
\frac{k}{k_L}
=
\pm
\arcsin
\frac{2}{\pi}
\sqrt
{
\frac
{|P|^2 n}
{
  2 J
  \left(
      U - 3 |P|
  \right)
}
}
\;.
\end{equation}

\begin{figure}[t]
\centering

\psfrag{p}[c]{$(+)$}
\psfrag{m}[c]{$(-)$}

  \hspace{1.5cm}\includegraphics[width=3.cm]{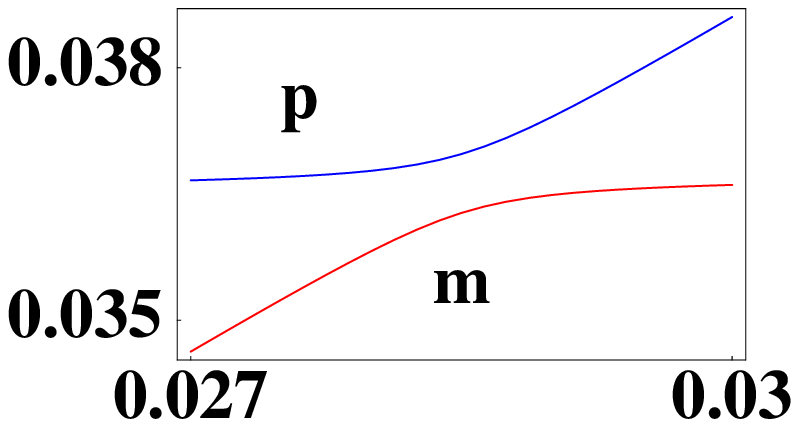}

  \vspace{-1.8cm}

\psfrag{o}[b]{\rotatebox{180}{$\hbar\omega_k/U_\Lambda$}}
\psfrag{k}[t]{$k/k_L$}

\psfrag{p}[c]{$(+)$}
\psfrag{m}[c]{$(-)$}

  \includegraphics[width=7.cm]{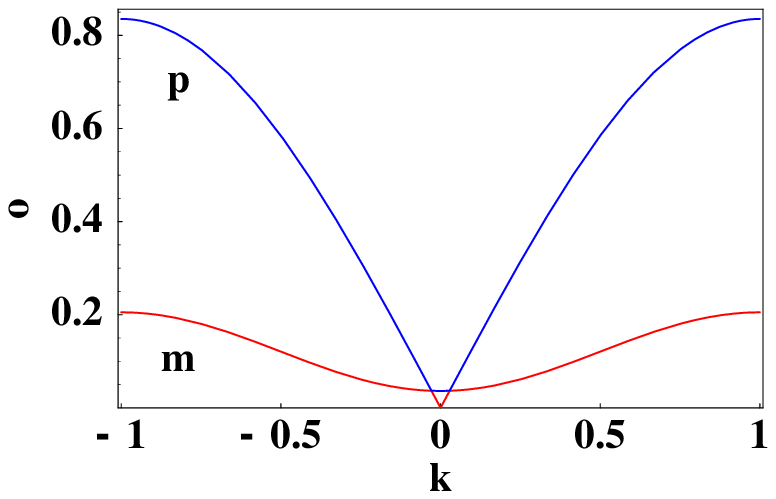}

\vspace{3mm}

\psfrag{c}[b]{\rotatebox{180}{$c_\Lambda/c_0$}}

  \includegraphics[width=6cm]{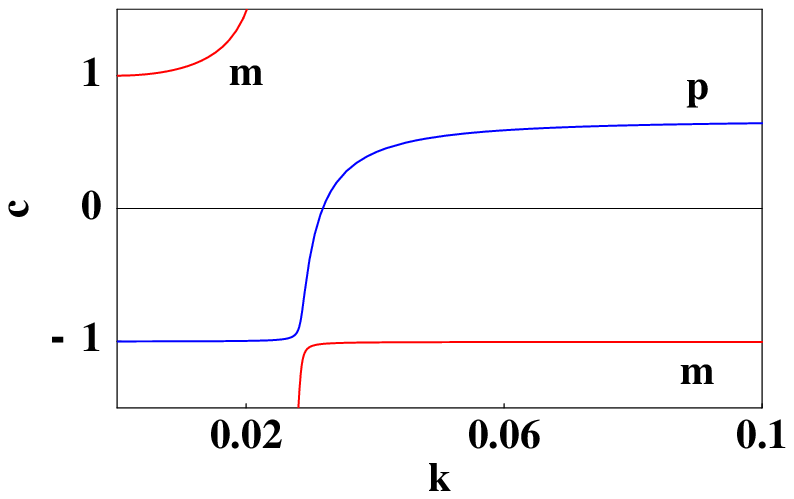}

\caption{Spectrum of collective excitations $\hbar\omega_k$ and
         the ratio $c_{\Lambda k}/c_{0 k}$ for $^{87}$Rb ($g_a<0$).
         $q=\tilde\Omega_0^2/4\omega_R\Delta=-4.5$, $\theta=0.7^\circ$, $n=2$.
        }
\label{spectrum}
\end{figure}

In the case $\theta \ne 0$, there are also two excitation modes.
Let us consider the case $g_a<0$ first. Typical $k$ dependences
for $n_\Lambda \ne 0$, $n_0 \ne 0$, are shown in
Fig.~\ref{spectrum}. The two branches do not cross anymore and the
coefficients $c_{\Lambda k}$, $c_{0 k}$ depend strongly on $k$
near the points of avoided crossings~(\ref{cp}), where the type of
oscillations changes. If we increase $\theta$ higher then a
certain value the condition~(\ref{ineq}) is not fulfilled.
According to Eq.~(\ref{n0}), Eqs.~(\ref{Delta}) become decoupled
and the energies of the eigenmodes are given by
\begin{eqnarray}
\label{null}
\hbar\omega_{0k}
&=&
\sqrt
{
  {\cal E}_0(k)
  \left(
      {\cal E}_0(k)
      +
      2 U_0 n
  \right)
}
\;,
\\
\hbar\omega_{\Lambda k}
&=&
\left\{
 \left[
     {\cal E}_\Lambda(k)
     -
     \left(
         U_0 - K
     \right)
     n
     +
     2
     \left(
     J_0
     -
         J_\Lambda
     \right)
     +
     \delta
 \right]^2
\right.
\nonumber\\
\label{lambda}
&&
\left.
 -
 (P n)^2
\right\}^{1/2}
\;,
\end{eqnarray}
where ${\cal E}_{0(\Lambda)}$ is given by Eq.~(\ref{ld}) with $J$
replaced by $J_{0(\Lambda)}$. Since inequality (\ref{ineq}) does
not hold and ${\cal E}_{\Lambda}(0)=0$, the $\Lambda$ mode has a
gap at $k=0$. Moreover, $U_0-K$ ($\approx |P|$) is positive as
well as all other summands in the square brackets in
Eq.~(\ref{lambda}). This means that for some nonvanishing values
of $k$ and large enough $n$ the frequency $\omega_{\Lambda k}$ can
become imaginary, i.e., the system can be unstable.

In the case $g_a>0$, $\theta \ne 0$, the excitation modes are
given by Eqs.~(\ref{null})~and~(\ref{lambda}), and $K-U_0$
($\approx P$) is positive. Therefore the frequencies
$\omega_{\Lambda k}$ are always positive and there is a gap at
$k=0$. For $\theta=0$ this spectrum reduces to that given by
Eq.~(\ref{Gold}) as it should.

\section{Conclusions}

We have investigated QPT of spin-1 bosons in an optical lattice
created by the lin-$\theta$-lin laser configuration. The laser
beams create not only periodic potentials for atoms, but also
couple the atomic ground states with magnetic quantum numbers
$m=\pm 1$. As a result there are only two lowest-energy Bloch
bands ($0$ and $\Lambda$) which should be taken into account and
the original three-component system is reduced to a two-component
one.

The interaction between the atoms is characterized by two
scattering lengths, the symmetric $g_s$ and the much smaller but
very important asymmetric $g_a$. This setup could be realized with
Rb or Na atoms, for which $g_a$ is negative (ferromagnetic
interaction) and positive (antiferromagnetic interaction),
respectively. It turned out that an appropriate change of the
easily tunable parameters of the setup, the laser intensity, and
the angle $\theta$, permits a very rich scenario of phase
transitions which we explored in detail in Section IV. The
symmetries of the Hamiltonian are discussed and it is shown that
the properties are essentially different for the atoms with
ferromagnetic and antiferromagnetic interactions. Depending on the
sign of the asymmetric scattering length, the isospin symmetry is
or is not spontaneously broken in the superfluid phase. The
corresponding collective modes, besides the always present
Bogoliubov mode, are isospin waves and either gapless or gapped at
$k=0$. A surprising feature we found in this study is that the
phase transition between Mott and superfluid phase can also be of
first order in certain domains of parameter space, with associated
domains where both phases can coexist, but, as we found, this is
only possible for $g_a>0$.

\begin{acknowledgments}
This work has been supported by the SFB/TR 12
``Symmetries and universalities in mesoscopic physics".
\end{acknowledgments}

\appendix*

\section{Coefficients in Eq.~(\ref{qwe})}

The coefficients in Eq.~(\ref{qwe}) are given by
\begin{eqnarray}
A_{\Lambda\Lambda}^{(1)}
&=&
- 2 J_\Lambda
\left(
    2 U_\Lambda n_\Lambda
    +
    K n_0
    -
    \mu
\right)
\\
%
A_{\Lambda 0}^{(1)}
&=&
2
\left[
    J_0
    \left|
        P
    \right|
    -
    J_\Lambda
    \left(
        K
    -
    \left|
        P
    \right|
    \right)
\right]
n_\Lambda
\\
%
A_{\Lambda\Lambda}^{(0)}
&=&
2 J_\Lambda^2
-
4 |P|
\left(
    K - |P|
\right)
n_0 n_\Lambda
\nonumber\\
&&
+
\left[
    U_\Lambda n_\Lambda
    +
    \left(
        K + |P|
    \right)
    n_0
    -
    \mu
\right]
\nonumber\\
&&
\times
\left[
    3 U_\Lambda n_\Lambda
    +
    \left(
        K - |P|
    \right)
    n_0
    -
    \mu
\right]
\\
%
A_{\Lambda 0}^{(0)}
&=&
2 n_\Lambda
\left\{
    \left[
        U_\Lambda n_\Lambda
    +
    \left(
        K + |P|
    \right)
    n_0
    -
    \mu
    \right]
    \left(
        K - |P|
    \right)
\right.
\nonumber\\
&&
\left.
    -
    |P|
    \left[
        3 U_0 n_0
    +
    \left(
        K - |P|
    \right)
    n_\Lambda
    -
    (\mu+\delta)
    \right]
\right\}
\end{eqnarray}
Expressions for $A_{00}^{(1)}$, $A_{0\Lambda}^{(1)}$, $A_{00}^{(0)}$,
and $A_{0\Lambda}^{(0)}$ can be obtained from these equations making the replacement
$\Lambda \to 0$, $0 \to \Lambda$, $\mu \to (\mu+\delta)$, $(\mu+\delta) \to \mu$.


\end{document}